\documentclass[10pt]{article}
\usepackage [cp1251] {inputenc}
\usepackage{amsfonts,amsmath,amsxtra,graphicx,amssymb,amscd}

\newcommand{\beq}{\begin{equation}}
\newcommand{\eeq}{\end{equation}}
\newcommand\bqa {\begin{eqnarray}}
\newcommand\eqa {\end{eqnarray}}
\newcommand{\bec}{\begin{cases}}
\newcommand{\eec}{\end{cases}}
\newcommand{\bei}{\begin{itemize}}
\newcommand{\eei}{\end{itemize}}
\newcommand{\bee}{\begin{enumerate}}
\newcommand{\eee}{\end{enumerate}}

\newcommand\pr {\partial}
\newcommand{\fr}{\frac}

\newcommand\nn {\nonumber}

\newcommand{\bear}{\begin{array}}
\newcommand{\enar}{\end{array}}

\newcommand{\braket}[1]{\Bigl\langle #1 \Bigr\rangle}

\newcommand{\sbraket}[1]{\left\langle #1 \right\rangle}

\begin{document}

\def\I{{\rm i}}

\def\h{\hbar}

\def\t{\theta}
\def\T{\Theta}
\def\w{\omega}
\def\ov{\overline}
\def\a{\alpha}
\def\b{\beta}
\def\g{\gamma}
\def\s{\sigma}
\def\l{\lambda}
\def\wt{\widetilde}
\def\t{\tilde}

\def\sumprime{\sideset\and^'\to\sum}

\vspace{-6.0cm}

\begin{center}
\hfill ITEP/TH-4/11\\
\end{center}

\vspace{1cm}

\centerline{\bf \Large Casimir effect for massless minimally coupled scalar field}
\centerline{\bf \Large between parallel plates in de Sitter spacetime}

\vspace{10mm}

\centerline{\bf Philipp Burda\footnote{burda@itep.ru}}

\vspace{5mm}

\centerline{ITEP, B. Cheremushkinskaya, 25, Moscow, Russia 117218}

\centerline{and}

\centerline{MIPT, Dolgoprudny, Russia}

\date{\today}

\begin{abstract}
Casimir effect for massless minimally coupled scalar field is studied. An explicit answer for de Sitter spacetime is obtained and analized. Cosmological implications of the result are discussed.
\end{abstract}


\section{Introduction}

Nowadays there is a rebirth of interest to the QFT on de Sitter background. Such a comeback is related to the recently discovered interesting features of interacting quantum fields in de Sitter spacetime, such as adiabatic catastrophe \cite{Pol-dSEternity}, UV/IR mixing and long time memory of correlations \cite{Pol-Krotov}, non-unitarity of such theories \cite{Emil-Buiv1} and new kind of IR divergencies \cite{confPI}. In the present paper we study free fields in de Sitter universe, but with imposed boundary conditions. This could be interesting at least, because sometimes special configurations for free fields could mimic behavior of a theory with interaction (see e.g. the case of the free field in contracting universe in \cite{Pol-Krotov}).

Usual motivation for such a study is the hope for the resolution of the cosmological constant problem by calculation of back reaction on the Gibbons--Hawking pair creation in de Sitter spacetime. In our previous work \cite{Emil-Burda} the program of breaking undesirable symmetry of the system by imposing appropriate boundary conditions for back reaction type problems was initiated and turned out to be successful for accounting for the back reaction on electromagnetic pair creation. Although our work was criticized, we believe that proposed procedure of ``cutting'' spacetime is relevant and helpful for the back reaction type of problems.

On the next step, when our strategy is applied to the theory of gravity, one arrives to the essential and famous for a long time difficulty, called {\it Casimir effect}. The point is that for gravity the role of back reaction source is played by v.e.v. of stress-energy tensor $\sbraket{T_{\mu\nu}}$, because it is a current which is usually situated on the r.h.s. of the equations of motion. Whereas in QED the source of back reaction is a v.e.v. of electromagnetic current $\sbraket{j_{\mu}}$. Therefore, if some boundary conditions are imposed on quantum field, in case of gravity even in the flat space one observes non trivial and physically testable consequences ($\sbraket{T_{\mu\nu}}_{b.c.} \neq 0$ -- this is exactly what Casimir has discovered). But in the case of QED, at least in flat space without any backgrounds, boundary conditions do not produce any new physics ($\sbraket{j_{\mu}}_{b.c.} = 0$).

Thus, if we want to try to develop our approach and consider back reaction on gravitational pairs creation, we need to understand better the influence of boundaries on v.e.v. of stress tensor, in order to find a possibility to extract from it a part interesting for back reaction.

It appears that the Casimir effect on curved backgrounds, such as de Sitter space, has not been studied so well as in the flat space. There are only a few works \cite{Sah1,Setare1}, where the Casimir effect between two parallel conducting plates in de Sitter spacetime was discussed. We mainly concentrate on this example as well, but with one important difference --- we consider the case of massless and minimally coupled scalar field, which was not considered in the mentioned papers, but is, in fact, important for the modern developments of QFT on de Sitter background due to its peculiar IR behavior. We use the approach of Brown and Maclay \cite{Brown-Maclay} with some small modifications along with its extension to the case of scalar field and curved spacetime. In order to explain and justify this approach we begin with detailed consideration of the standard case of flat spacetime and then proceed to the de Sitter space. It is probably worth pointing out here, that even in the case of Minkowski spacetime our approach illuminated some details, which were hidden in previous considerations or at least discussed in much less details.

The work is organized as follows, in the second section we explain our strategy in details. Beginning from the third section we consider concrete examples and start from Minkowski spacetime. In the fourth section we examine the same setting, but in de Sitter space, and we also discuss cosmological implications of our result at the end of the paper. We work in the system of units where $\h=c=G=1$ and the signature of the metric is $(+,-,-,-)$.

\section{Method of calculation}

In this paper we consider the classical formulation of the Casimir effect. We study the renormalized v.e.v. of the stress-energy tensor for massless minimally coupled scalar field between two perfectly conducting parallel planes, situated orthogonally to the $z$ axis at $z=0$ and $z=L$. The action
\beq
S = \int d^4x \left\{ \sqrt{-g(x)} g^{\mu\nu}(x) \pr_{\mu}\phi\pr_{\nu}\phi \right\} ,
\eeq
gives rise to the following expression for the stress-energy tensor
\beq
T_{\mu\nu} = \pr_{\mu}\phi\pr_{\nu}\phi - \fr{g_{\mu\nu}}{2} \pr_{\rho}\phi\pr^{\rho}\phi = \left[ \left\{ \pr_{x^\mu}\pr_{y^\nu} - \fr{g_{\mu\nu}}{2} \pr_{x^\rho}\pr^{y_\rho} \right\} \phi(x)\phi(y) \right]_{y=x}
\eeq
Therefore, in order to calculate its v.e.v. one could just act by the specified differential operator on the appropriate Green function (which in our case should satisfy Direchlet boundary conditions at $z=0$ and $z=L$ --- $\phi_{z=0} = \phi_{z=L}=0$) and then take the limit of coincident points.
\beq
\braket{T_{\mu\nu}} = \left[ \left\{ \pr_{x^\mu}\pr_{y^\nu} - \fr{g_{\mu\nu}}{2} \pr_{x^\rho}\pr^{y_\rho} \right\} \braket{\phi(x)\phi(y)}_{D(0,L)} \right]_{y=x} = \left[ \mathcal{T}_{\mu\nu} \cdot \braket{\phi(x)\phi(y)}_{D(0,L)} \right]_{y=x}
\eeq
To get rid of additional and not interesting divergencies it is convenient to use, so-called, symmetric two-point function $G^{(1)}(x,y)$, which by definition is equal to the v.e.v. of anti-commutator of the fields
\beq
G^{(1)}(x,y) = \braket{\left\{\phi(x),\phi(y)\right\}} = \braket{\phi(x)\phi(y) + \phi(y)\phi(x)}
\eeq
In the limit of coincident points it gives only the additional factor of $2$
\beq
\left. \braket{\phi(x)\phi(y)} \right|_{y=x} = \fr{1}{2} \left. G^{(1)}(x,y) \right|_{y=x}
\eeq
but the pleasant thing is that expression for $G^{(1)}(x,y)$ is simpler than for $\sbraket{\phi(x)\phi(y)}$.

Thus, we have the following expression for the v.e.v. of the stress-energy tensor:
\beq
\braket{T_{\mu\nu}} = \fr{1}{2} \left[ \mathcal{T}_{\mu\nu} \cdot G^{(1)}(x,y)_{D(0,L)} \right]_{y=x}
\label{vev2}
\eeq
The problem is reduced now to the construction of the Green function $G^{(1)}(x,y)_{D(0,L)}$ for the theory with two perfectly conducting plates at $z=0$ and $z=L$. Such a Green function could be obtained from the empty space Green function $G^{(1)}(x,y)$, using the well known high--school trick, called {\it method of images}. One only needs to add an infinite number of mirror sources, as shown and explained on the Figure~1, and then the boundary conditions will be automatically satisfied.

\begin{figure}[h]
\begin{center}
\includegraphics[scale=0.7]{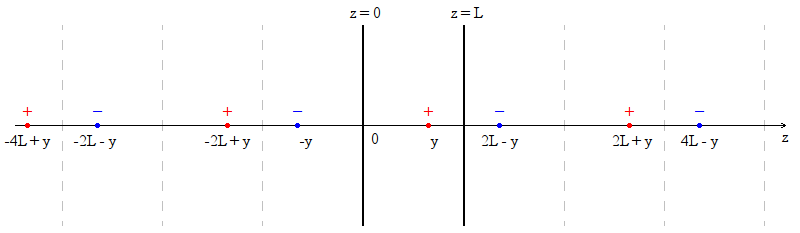}
\end{center}
\label{ms-moi}
\caption{\footnotesize Positions of mirror sources. Building method: starting from real source with positive ``charge'' at the point $y$, one needs to built its images in the ``mirrors'' at $z=0$ and $z=L$, but already with negative charge, and then, because each charge (real or image) should be reflected in each mirror, one should reflect left image ($-y$) in the right mirror, and right image ($2L-y$) -- in the left one. In this way one obtains second generation of images (they again have positive ``charge''), which in turn produce third generation of images and so on.}
\end{figure}

\beq
G^{(1)}(x,y)_{D(0,L)} = \sum_{n=-\infty}^{\infty} \left[ G^{(1)}(x,y^+_n) - G^{(1)}(x,y^-_n) \right], \quad\quad (y^{\pm}_n)_z = \pm 2nL \pm y_z
\eeq
Substituting this expansion into (\ref{vev2}) and changing the order of the action of the operator $\mathcal{T}_{\mu\nu}$ and the summation
\beq
\braket{T_{\mu\nu}} = \fr{1}{2} \sum_{n=-\infty}^{\infty} \left[ \mathcal{T}_{\mu\nu} \cdot G^{(1)}(x,y^+_n) - \mathcal{T}_{\mu\nu} \cdot G^{(1)}(x,y^-_n) \right]_{y=x} = \fr{1}{2} \sum_{n=-\infty}^{\infty} \biggl[ t^{+}_{\mu\nu}(x,n) - t^{-}_{\mu\nu}(x,n) \biggr]\label{fil}
\eeq
we arrive at the almost final desirable expression for $\sbraket{T_{\mu\nu}}$, the only its disadvantage is that this expression is divergent. The most straightforward and simple method of curing out such a divergence, based on subtraction of an empty space part of the v.e.v, works perfectly well also in the de Sitter spacetime (because of its high symmetry). Hence, we apply this, standard for Minkowski spacetime, subtraction scheme
\beq
\braket{...}_{ren} = \braket{...}_{with\,b.c.} - \braket{...}_{empty\,space}.
\eeq
In our notations this regularization procedure is equivalent to the subtraction of the term $t^{+}_{\mu\nu}(x,0)$ with the source at the point $y^{+}_{0}$ from (\ref{fil}). Hence, finally we obtain
\beq
\braket{T_{\mu\nu}}_{ren} = \fr{1}{2} \Biggl[ \sum_{n=-\infty}^{\infty}{}^{\hspace{-0.21cm}\prime} \;\, t^{+}_{\mu\nu}(x,n) - \sum_{n=-\infty}^{\infty} t^{-}_{\mu\nu}(x,n) \Biggr]
\eeq
Every component of this expression can be easily calculated, but for Minkowski and de Sitter spacetimes one can also sum them up and obtain an explicit answer for $\sbraket{T_{\mu\nu}}$.

For the sake of brevity in all the examples we will consider not the stress tensor itself, but the energy per unit area of the field between the plates. Expression for the energy can be easily obtained from $T_{00}$ component of the stress tensor by the simple integration over $z$
\bqa
E & = & \int\limits_0^L dz \; \braket{T_{00}}_{ren} = \fr{1}{2} \int\limits_0^L dz \Biggl[ \sum_{n=-\infty}^{\infty}{}^{\hspace{-0.21cm}\prime} \;\, t^{+}_{00}(x,n) - \sum_{n=-\infty}^{\infty} t^{-}_{00}(x,n) \Biggr] = \nn\\
& = & \fr{1}{2} \left[ \sum_{n=-\infty}^{\infty}{}^{\hspace{-0.21cm}\prime} \;\, \int\limits_0^L t^{+}_{00}(x,n) dz -  \sum_{n=-\infty}^{\infty} \int\limits_0^L t^{-}_{00}(x,n) dz \right] = \fr{1}{2} \left[ \sum_{n=-\infty}^{\infty}{}^{\hspace{-0.21cm}\prime} \;\, \varepsilon^+_n -  \sum_{n=-\infty}^{\infty} \varepsilon^-_n \right]
\label{energy}
\eqa
Such an interchange of the summation and integration has also some physical meaning, because as we will show properly treated contribution of all negative charges ($y^-_n$) to the energy of the field vanishes. Whereas, their contribution to the energy density is some non-trivial function of $z$, which is divergent at the boundaries. Of course, there are also divergencies in some of $\varepsilon^-_n$, but there is an obvious way to handle them without any complication of the problem.

\section{Minkowski spacetime}

In this section we apply our general strategy, developed in the previous section, to the Minkowski spacetime to check its consistency and discuss everything in details on the concrete example.

Using expression for the symmetric function in empty and flat spacetime
\beq
G^{(1)}(x,y) = - \fr{1}{2\pi^2} \fr{1}{\left(t_x-t_y\right)^2-\left(\vec{x}-\vec{y}\right)^2}
\eeq
after some trivial calculations we obtain
\beq
\varepsilon^+_n = - \fr{1}{16\pi^2L^3} \fr{1}{n^4}
\eeq
As we expect, expression for $\varepsilon^+_0$ diverges. Nature of this divergence is clear, because, by definition, $\varepsilon^+_0$  corresponds to the energy in empty space without any boundaries. Character of the divergence ($\varepsilon^+_0 \sim \a^{-4}, \; \a \rightarrow 0$) also supports this observation.

For $\varepsilon^-_n$ we get
\beq
\varepsilon^-_n = \fr{1}{24\pi^2L^3} \fr{3n^2+3n+1}{n^3\left(n+1\right)^3}
\eeq
This expression is also divergent for $n = 0$ and $n = -1$. But it is worth stressing here that the sum $\varepsilon^-_{-1} + \varepsilon^-_{0}$ could be made finite, if one approaches divergencies simultaneously. From the fact, that Casimir effect is experimentally confirmed, we conclude that we act properly and these divergencies should cancel each other. Nevertheless, it is interesting to understand their nature. Notice, that $\varepsilon^-_{-1}$ and $\varepsilon^-_{0}$ correspond to the sources at points $2L-y$ and $-y$ respectively. In our notation, this is the first generation of images, i.e. they were obtained by reflecting ``real'' source in the mirrors. It is quite interesting, that divergent expressions arise only when one builds an image of original source in the mirrors of the systems. The same holds for de Sitter spacetime and again divergencies cancel each other. It seems, that such divergencies always arise in the presence of any boundaries and one, probably, should search their roots in {\it Heisenberg's uncertainty principle} --- in the problem in question one tries to specify values of the filed on the boundary, hence momentum of the field (stress-energy tensor) diverges. As far as we know, such picture was proposed in \cite{Ford-Svaiter}. According to our results it is also clear, that some smearing procedure should be applied to each boundary, when they are considered separately. But as we have seen, these divergencies eliminate each other, if we consider both boundaries simultaneously. Therefore, if we can think that we understand the nature of these divergencies, at least qualitatively, the question about their cancelation still remains. Why does such cancelation occur and does it happen in any spacetime, for any geometry and for any external conditions (temperature, presence of matter)?

Let's now return to the Casimir effect for Minkowski spacetime, and, using expressions for $\varepsilon^+_n$ and $\varepsilon^-_n$, we get well-known result for the Casimir energy between two plates at distance $L$
\beq
E_{M} = - \fr{1}{1440} \fr{\pi^2}{L^3}
\eeq

As we have mentioned at the end of the previous section, it appears that properly treated sum of all negative mirror sources gives zero contribution to the final answer for the energy, i.e.
\beq
\sum_{n=-\infty}^{\infty} \varepsilon^-_n = 0
\eeq
The same is true for de Sitter as well. This fact should have some trivial explanation, but at the moment we cannot provide it. There is quite simple mathematical hint and some speculations about electromagnetic field in \cite{Albe-Far-Theo}, but we actually seek for some physical reason for such a vanishing in the case of massless scalar field.

\section{de Sitter spacetime}

In this section we proceed with the consideration of the Casimir effect for two parallel plates in de Sitter spacetime. But, at first, we make some comments about the geometry of de Sitter spacetime and say a few words about QFT on such a background \cite{BD}.

de Sitter spacetime is a vacuum solution of Einstein equations with the positive cosmological constant. It is convenient to consider it as a hypersurface in a flat five-dimensional space $\eta_{MN} = \mbox{diag}(1,-1,-1,-1,-1)$, satisfying the following equation
\beq
X^MX^N \eta_{MN} = - \fr{1}{H^2}.
\eeq
Introducing different parametrizations of this hypersurface one could obtain different metrics for de Sitter. Convenient for our purposes coordinate system (usually referred to as planar) is specified by the following parametrization
\bqa
X^0 & = & \fr{\sinh(Ht)}{H} + \fr{H}{2} e^{Ht} \left|\vec{x}\right|^2\nn\\
X^i & = & x^i e^{Ht}, \quad i=1,2,3 \nn\\
X^4 & = & \fr{\cosh(Ht)}{H} - \fr{H}{2} e^{Ht} \left|\vec{x}\right|^2, \quad -\infty < t,x_i < \infty
\eqa
It covers only half of the hyperboloid, but this is not a problem, because it can be resolved by further transformations. The line element in the above coordinates  has a quite simple form $ds^2 = dt^2 - e^{2Ht} \left(dx_1^2 + dx_2^2 + dx_3^2 \right)$, from which it is easy to observe the expanding nature of the de Sitter universe. Making a further transformation and proceeding from, so called, cosmic time $t$ to the conformal time $\tau = - \fr{1}{H} e^{-Ht}$ ($-\infty < \tau < 0$), one obtains conformal form of the de Sitter metric
\beq
ds^2 = \fr{1}{H^2\tau^2} \left( d\tau^2 - dx_1^2 - dx_2^2 - dx_3^2 \right)
\eeq
which we will use for calculations. Notice, that one could cover whole de Sitter space by expanding the range of $\tau$ to the whole real line.

The only thing we want to stress about QFT on de Sitter background, is that there is no de Sitter invariant vacuum state for the case of massless minimally coupled scalar field \cite{Allen}. The common strategy is to look for states, that break de Sitter invariance as little as possible \cite{Allen-Folacci}. We choose the $E(3)$ invariant vacuum, because it corresponds to the Bunch-Davies vacuum for the massive case, and also in such a vacuum $\sbraket{\phi^2} = const$ (i.e. does not depend on time) for flat spacelike surfaces, which matches with our geometry. For such a vacuum in the conformal coordinates function $G^{(1)}(x,y)$ reads \cite{Allen}
\beq
G^{(1)}(x,y) = \fr{H^2}{4\pi^2} \left( \fr{2\tau_x\tau_y}{\left(\vec{x}-\vec{y}\right)^2-\left(\tau_x-\tau_y\right)^2} - \log\left[\left(\vec{x}-\vec{y}\right)^2-\left(\tau_x-\tau_y\right)^2 \right] \right)
\eeq
Computations for positive sources give
\beq
\varepsilon^+_n = \fr{H^2}{16} \fr{n^2L^2-\tau^2}{\pi^2L^3n^4} = \fr{H^2}{16\pi^2L} \fr{1}{n^2} - \fr{H^2\tau^2}{16\pi^2L^3} \fr{1}{n^4}
\eeq
One can see that two type of divergencies at $n = 0$ are present here. The first term appears because we consider non-conformal coupling of scalar field. It has special for de Sitter spacetime divergency, which is related to the fact, that chosen vacuum breaks the de Sitter invariance. This term also has unusual for Casimir energy dependence from the distance between the plates $\sim L^{-1}$. The second term has standard Minkowskian divergency and, in fact, it is equal to the $\varepsilon^+_n$ for Minkowski spacetime multiplied by the expansion factor $H^2\tau^2$. Notice, that the first term has an opposite sign, hence it compensates contribution of the second term. Its presence is the manifestation of antigravitational nature of $\Lambda$ in de Sitter universe.

Contribution of the negative charges
\beq
\varepsilon^+_n = \fr{H^2}{48\pi^2L^3} \fr{9n^2L^2\left(n+1\right)^2 + 2\tau^2(3n^2+3n+1)}{n^3(n+1)^3} = \fr{3H^2}{16\pi^2L} \fr{1}{n(n+1)} + \fr{H^2\tau^2}{24\pi^2L^3} \fr{3n^2+3n+1}{n^3(n+1)^3}
\eeq
also has two analogous terms. They both diverge at $n=-1$ and $n=0$ and for both of them these divergencies cancel each other.

Finally, using (\ref{energy}) for the Casimir energy between two plates at distance $L$ in de Siter spacetime, we obtain
\beq
E_{dS} = \fr{H^2}{1440} \fr{15L^2-\pi^2\tau^2}{L^3} = \fr{H^2}{96L} - \fr{H^2\tau^2\pi^2}{1440L^3}
\eeq
Returning to the cosmic time $t$ and expressing everything through the physically measurable distance in the expanding unverse
\beq
L_{phys} = \int \sqrt{\g_{\a\b}dx^{\a}dx^{\b}} = \int \sqrt{-g_{\a\b}dx^{\a}dx^{\b}} = \int\limits_0^L \sqrt{-g_{zz}} \; dz = - \fr{L}{H\tau} = L e^{Ht}
\eeq
one could rewrite the expression for the Casimir energy in physically meaningful terms
\beq
\boxed{
E_{dS} = \left( \fr{H^2}{96L_{phys}} - \fr{\pi^2}{1440L^3_{phys}} \right) e^{Ht}
}
\label{answer}
\eeq
Observing the structure of this answer one should pay attention to the presence of $H^2$ factor, which is related to cosmological constant $\Lambda$ and scalar curvature $R$ ($R = 4 \Lambda = 12 H^2$). First of all, the presence of $H^2$ in the first term provides that in a flat space limit ($H \rightarrow 0$) the answer (\ref{answer}) reproduce the result for the Minkowski spacetime $E_{ds} \rightarrow E_{M}, \, H \rightarrow 0$. But in highly curved de Sitter universe, conducting plates repulse from each other ($E_{dS} > 0$) for most of the distances. It is easy to see this, from the dependence of $E_{dS}$ on $L_{phys}$ for different values of $H$, which is present at Figure~2. The critical length $L_{cr}$ ($E_{ds}(L_{cr}) = 0$), at which neither attraction not repulsion is present, equals to $L_{cr} = \fr{\pi}{\sqrt{15}H}$, and tends to $0$ when $H$ is growing.
\begin{figure}[h]
\begin{center}
\includegraphics[scale=0.5]{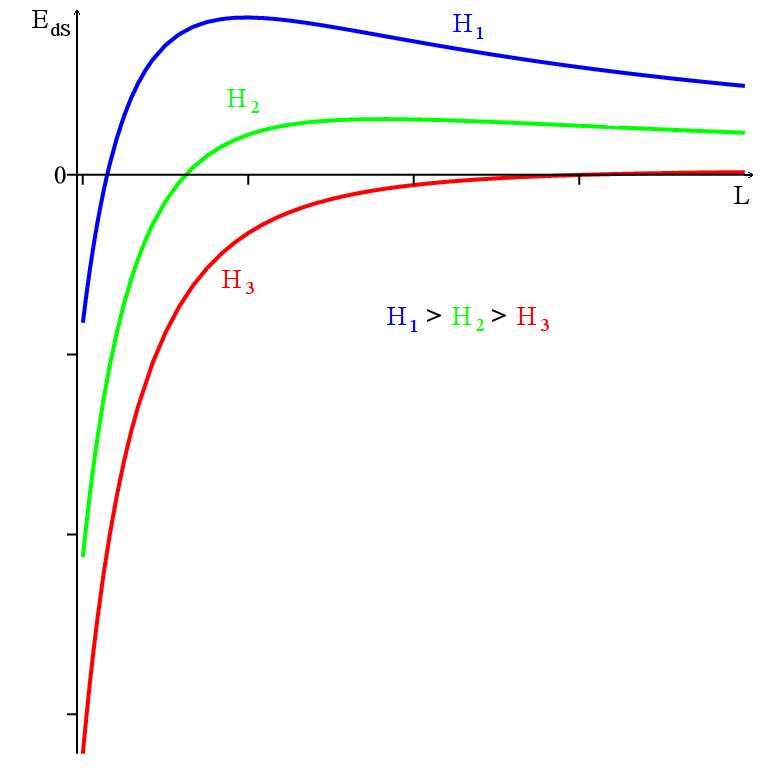}
\end{center}
\label{EdS}
\caption{\footnotesize Casimir energy for massless minimally coupled scalar field between two plates at distance $L$ in de Sitter spacetime, as a function of $L_{phys}$, for different values of $H$.}
\end{figure}

Cosmological implications of this result differ from artificial situation with arbitrary large and curved de Sitter spacetime discussed above. At the early stages of evolution of the Universe, when, as it is believed, cosmological constant was significant the Universe itself was quite small, because the cosmological event horizon was of the size $L_{hor} = \fr{1}{H}$ \cite{Bousso}. Thus, $\fr{L_{hor}}{L_{cr}} = \fr{\pi}{\sqrt{15}} \approx 0.81$, which means that at all the times during the evolution of the Universe plates at distances smaller than $ \sim 80\%$ of radius of the Unverse feel an attractive force and only plates at huge distances comparable with the radius of the Universe are repulse from each other (this is the case of $H_3$ (red line) at the Figure~2). Of course, all these conclusions are valid only for the toy model of an empty de Sitter universe with negligible pairs production rate.

\section{Conclusions}

In this paper we have analyzed the Casimir energy between two conducting plates in the de Sitter spacetime for the massless minimally coupled scalar field. The method with mirror sources was generalized and applied to the curved background. Implication of this method allowed us to point out some interesting aspects of divergencies, common for both Minkowski and de Sitter spacetimes. There are even some moments, for which we do not have yet a clear physical explanation at present. But, Casimir effect, especially for flat spacetime, is so well understood, that most likely we looked not good enough.

Justification for writing such a simple paper was an absence of the consideration of the massless and {\it minimally coupled} scalar field in de Sitter spacetime in the literature (at least, we didn't find it). Meanwhile, such a case is important for the issue of the unusual IR behavior of QFT on de Sitter background. Therefore, for us the main advantage of the mirror sources method was the fact that it allowed us to obtain the exact answer for the Casimir energy in the explicit form (\ref{answer}). The answer has a remarkable structure with two competitive terms, negative one having the same nature as in Minkowski spacetime, and positive one, special for de Sitter. The last one has the unusual dependence on the distance between the plates and is proportional to the scalar curvature of the universe. Which allows us to speculate about cosmological aspects of our result.

We consider the result of the present paper just as a first very simple step. All the results can be generalized in many different directions, beginning from the other geometries of the problem (concentric spheres, pistons etc.), different external conditions (thermal bath, presence of matter etc.), another field content --- electromagnetic and massive fields (this case is especially important for back reaction side of the problem) and finally proceeding to the arbitrary FRW or even more general spacetimes.

\section{Acknowledgments}

I would like to thank E.Akhmedov for initiation of this work and very valuable discussions. I also would like to acknowledge V.Serbinenko for the help with the text and support. This work is partly supported by RFBR grant 11-02-01220, by joint grant 11-01-90436-Ukr, by the Dynasty Foundation and by Federal Agency for Science and Innovations of Russian Federation (contract 14.740.11.0081).

\end{document}